\documentclass{article}[12pt]
\usepackage[utf8]{inputenc}
\usepackage[english]{babel}
\usepackage{xcolor}
\usepackage{caption}
\usepackage{subcaption}
\usepackage[font={footnotesize, bf}]{caption}
\usepackage[font={footnotesize, bf}]{subcaption}
\usepackage{graphicx}
\graphicspath{ {images/Pictures} }
\usepackage{svg}
\usepackage{amsmath}
\usepackage{amssymb}
\usepackage{float}
\usepackage{hyperref}
\usepackage{geometry}
\usepackage{booktabs}
\usepackage{bm}
\usepackage{enumitem}
\usepackage{tikz}
\usepackage{siunitx}
\usepackage{physics}
\usepackage{multicol}
\usepackage{lipsum}
\usepackage{subcaption}
\usepackage{fancyhdr}
\usepackage{nomencl}
\usepackage{blindtext}
\usepackage[dvipsnames]{xcolor}
\usepackage{pdfpages}
\usetikzlibrary{calc,patterns,decorations.pathmorphing,decorations.markings, intersections}
\usepackage{pgfplots}
\pgfplotsset{compat=1.11}
\usepgfplotslibrary{fillbetween}
\pgfdeclarelayer{bg}
\pgfsetlayers{bg,main}

\tikzset{photon/.style={
  decorate,
  decoration={snake, amplitude=0.5mm, segment length=4mm},
  draw=orange!75
}}

\usepackage[
    natbib=true,
    style=nature,
    sorting=none
]{biblatex}
\usepackage{xurl}

\usepackage{mathtools}

\usepackage{titlesec}
\titleformat*{\section}{\Large\bfseries}
\titleformat*{\subsection}{\normalsize\bfseries}
\titleformat*{\subsubsection}{\normalsize\bfseries}
\titleformat*{\paragraph}{\normalsize\bfseries}
\titleformat*{\subparagraph}{\normalsize\bfseries}

\begin{document}

\twocolumn[
\begin{center}
    \textbf{\Large{Scheme for Transport-based Global Entanglement Distribution using Quantum Processors}}
    
    \vspace{0.2cm}
    \small{Erik Lundblad, Mira Abu Yahia, Antonius Johannes Renders, Andreas Walther, Adam Kinos, Lars Rippe} \\
    \textit{\small{Department of Physics, Lund University, P.O. Box 118, SE-22100 Lund, Sweden}} \\
    \makeatletter
    \renewenvironment{abstract}{
        \if@twocolumn
          \section*{}
        \else
          \section*{}
        \fi}
        {\if@twocolumn\else\endquotation\fi}
    \makeatother
    \vspace{-8ex}
    \begin{abstract}
    We propose a scheme for distributing entanglement over global distances in a heralded manner by using satellites to physically transport entangled processor nodes with rare-earth-ion qubits. 
    A full analysis of channel losses, errors and background light is performed to determine the fidelity and number of entangled pairs that can be distributed between two ground stations. 
    We show that the scheme works already with a single satellite and can distribute close to the theoretical maximum number of entangled pairs that can be generated in a satellite overpass. In addition, we argue that in theory transportation-based schemes outperform other satellite-based schemes and can be scaled up to a constellation without additional channel losses. 
    Daytime operation seems feasible as long as the sky is clear, with an EPR pair fidelity ranging from $99.3\%$ at shorter network lengths to $93.9\%$ with global coverage and can be further improved by active error correction or entanglement purification. 
    \end{abstract}
\end{center}
\vspace{1ex}
 ]


\noindent Distributing entanglement over global distances shows promise in enabling several quantum technologies such as global synchronisation of atomic clocks \cite{Kmr2014,Ilo-Okeke2018,Jozsa2000,Yurtsever2002}, enhancing the performance of distributed quantum computing \cite{Cirac1999} and device authentication \cite{Li2007}. Similarly, techniques for increased measurement sensitivity within astronomy have been proposed, with suggestions for interferometry without transmission losses at arbitrarily long baselines \cite{Bland-Hawthorn2021,Khabiboulline2019_1,Khabiboulline2019_2,Gottesman2012} and with a resolution beyond the standard quantum limit \cite{Nichol2022,Guo2020}.

Currently, entanglement distribution is heavily limited by the intrinsic channel losses in quantum communication networks \cite{Pirandola2017}, rendering otherwise reliable systems such as optical fibers irrelevant on a global scale without extensive use of quantum repeaters every tens of kilometers \cite{Muralidharan2014,Duan2001}. To overcome these losses without repeaters, a higher source rate of photons need to be used in order to increase the overall distribution rate. However, packing photons closer in time increases the frequency bandwidth of each photon, meaning that the overall channel capacity remains unaffected. Channel losses thus set a fundamental limit to achievable distribution rates and finding global entanglement distribution schemes with lower losses will lead to higher channel capacities in the long run.

Several global communication schemes have therefore been proposed using free-space links in very low earth orbit (VLEO) where the transmission losses are lower \cite{Liu2026,Goswami2023,Gndoan2021,Liorni2021,Khatri2021,Boone2015}. However, these schemes still need multiple links to reach global distances due to line of sight requirements and they suffer from inter-link losses. To minimize the number of links, long-lived quantum memories can instead be utilised to physically transport quantum states onboard individual satellites \cite{Gndoan2024,Wittig2017}, where the network length is only limited by the coherence time of the memory. The only other losses in such a scheme stems from transmitting the information to and from the satellite, a loss that is common for all satellite schemes. This type of transportation-based schemes should therefore have the highest theoretically achievable channel capacity of any satellite scheme.

However, to the best of the authors' knowledge, no proposition for a highly multi-mode quantum memory has yet described a method for resetting individual modes of information without altering other modes. Any entanglement distribution scheme involving quantum memories will thus operate highly inefficiently as long as there are channel losses. Furthermore, no proposed scheme for distributing entanglement globally has shown compatibility with active error correction, fundamentally limiting current schemes to first generation quantum communication systems.

To combat these limitations, we propose a scheme for distributing entanglement globally in a heralded manner using quantum processor nodes made out of individual rare-earth-ions (REIs) as qubits. First, heralded entanglement generation (HEG) is used to create entangled Einstein-Podolsky-Rosen (EPR) pairs between qubits in a satellite and qubits in a ground station over a free-space channel. Crucially, if an HEG trial fails, the qubits used can be re-initialised and the HEG trial re-attempted until it succeeds. The satellite then travels to a different ground station where the process is repeated. When EPR pairs entangled to each ground station exist, deterministic entanglement swapping can be used to generate EPR pairs directly between the two ground stations. Thus, global distances can be reached already with a single satellite if the qubits have a coherence time longer than half of the orbital period of the satellite, which in VLEO is approximately 45 minutes. Even though our protocol is in principle compatible with different platforms, REIs are a strong candidate by being a solid-state system with shown coherence times greater than 13 hours in the hyperfine ground states \cite{Wang2025}.

With the use of qubits, this scheme is the first global entanglement distribution scheme compatible with active error correction, which in turn would allow for second (and potentially third) generation quantum communications systems in the long term \cite{Muralidharan2016}. In the short term, the possibility of re-attempting failed HEG trials until they succeed will lead to a vast improvement in how efficiently the information modes are being used and how many are needed.

\begin{figure*}[t!]
\scalebox{1}{\input{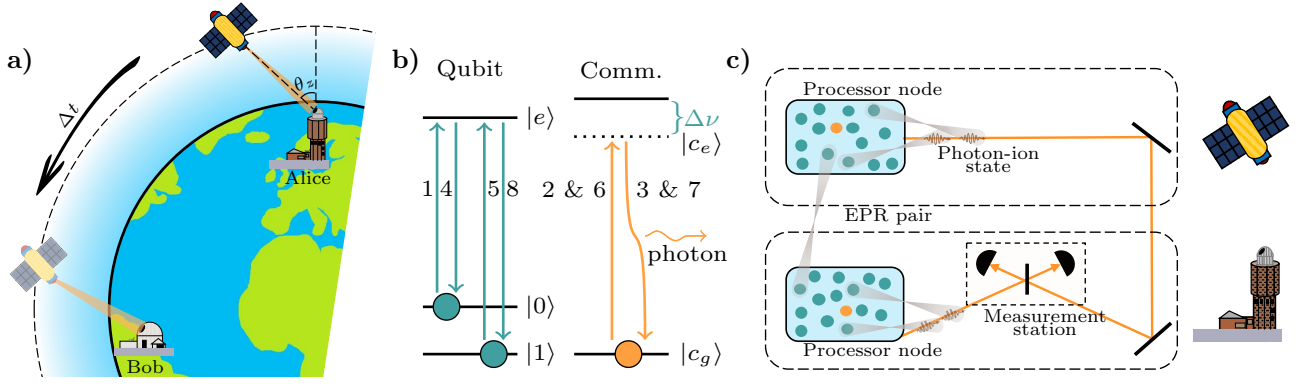}}

\caption{\textbf{a)} A satellite passes by ground station Alice, creating the zenith angle $\theta_{z}$ with regards to the ground station. During the overpass, a link of entangled EPR pairs is established. The satellite then travels a distance along its' orbital path and arrives at Bob, where it alternates between generating EPR pairs with Bob and performing entanglement swapping until all EPR pairs from Alice have been converted to EPR pairs between the two ground stations. 
\textbf{b)} A qubit is initialised in the $\frac{1}{\sqrt{2}}\qty(\ket{0}+\ket{1})$ state. If a qubit is excited into the $\ket{e}$ state, the communication ion is shifted an amount $\Delta\nu$ out of resonance through a Stark shift. Any attempt at driving the communication ion at its original frequency will then fail, allowing for a blockade gate. By following pulse sequence 1-8, a photon is emitted from the communication ion with the photon-ion state $\frac{1}{\sqrt{2}}\qty(\ket{0} a_L^\dagger + \ket{1} a_E^\dagger)$, where $a_E^\dagger$ and $a_L^\dagger$ denotes the creation of a photon in the early and late time-bin, respectively \cite{Kinos2025}. If an HEG trial fails, the qubits are reset and pulse scheme 1-8 is repeated to emit a new photon.
\textbf{c)} Bell-state measurement of photons from the satellite and a ground station is performed by overlapping their early and late time-bins respectively at a beam splitter before detection. One early and one late detected photon projects the photon-ion state on the EPR pair $\frac{1}{\sqrt{2}}\qty(\ket{01}\pm\ket{10})$ with one qubit in each node \cite{Kinos2025}.
}
\label{fig:entanglement_generation}
\end{figure*}

The article is divided as follows: the new scheme and the capacity of distributing EPR pairs is described. An estimation of the fidelity of the pairs is then made, taking into account channel losses and errors from initialisation, measurements, two-qubit gates, as well as background light. The theoretical maximal distribution rate of any satellite-based scheme is discussed and a comparison between them is made, where we argue that transportation-based schemes have a higher maximal channel capacity. Finally, the performance and scalability of a REI quantum processor node using feasible parameters is discussed.

\section*{Results}
\subsection*{Entanglement distribution scheme}

Each processor node will consist of two species of REIs that are randomly doped into a crystal host. In the first species, qubits are encoded into the ground-state hyperfine levels of individual REI. These ions can be addressed individually in frequency space with optical pulses, which allows us to perform single- and two-qubit gates using the dipole-dipole interactions between ions \cite{Kinos2021_2}. The second species act as communication ions, capable of emitting photons that are time-bin entangled to the qubit states \cite{Wesenberg2007, Kinos2025}, see Figure \ref{fig:entanglement_generation}. Each node consists of several qubit ions closely spaced in the crystal lattice, sharing one communication ion in their proximity. This allows for sequential readout of qubit states through the same channel.

To generate an EPR pair between the satellite and a ground station, a photon entangled to a qubit in the satellite is sent to the ground station, where it is collected by a telescope. A similar photon entangled with a ground station qubit is then temporally overlapped with the photon from the satellite by sending both photons through a beamsplitter. Through a Bell-state measurement of the two photons, an EPR pair with the two qubits can be generated and constitutes one HEG trial. The result of the HEG trial is then conveyed classically back to the satellite and if the HEG trial failed, the desired qubits can be re-initialised and reused in upcoming trials.

With multiple qubits available in the satellite node, HEG trials can be performed in quick succession by sending the photons in a pulse train towards the ground station and relaying the trial result to the satellite first after all available qubits have had a trial each. If the pulse train is longer than the roundtrip communication time between the satellite and the ground station, the trial rate is only limited by the speed at which photons can be emitted from the node. Consecutive rounds of HEG trials can then be made until a desired amount of EPR pairs have been generated with a ground station.


The probability $p_{\text{HEG}}$ of succeeding an HEG trial is given by:
\begin{align*}
    p_{\text{HEG}} 
    = \frac{1}{2}\eta_0^2\eta_c\qty(\theta_z)\mathcal{E}\qty(B)
    \label{eq:HEG_trial_success_probability}
    \tag{1}
\end{align*}
where $\eta_0$ is the channel-independent losses that combines the efficiency of emitting, collecting, and detecting a photon and $\eta_c\qty(\theta_z)$ is the channel efficiency of the free-space channel at an angle $\theta_z$ from the zenith. $\mathcal{E}(B)$ takes into account the effect of background light photons $B$ entering the HEG trial. Since background light affects the fidelity of the EPR pairs much earlier than $p_{\text{HEG}}$, $\mathcal{E}(B)$ is close to unity for all practical cases.
%
The channel efficiency and the effect of background light are described in detail in Methods \ref{appendix:Channel_Losses} \& \ref{appendix:Background_Light}, respectively.

To perform entanglement swapping, one EPR pair from each ground station is needed per swapped pair. Consequently, if the satellite flies from Alice to Bob, the maximal number of EPR pairs that can be generated between the ground stations is given by the number of EPR pairs generated over Alice. By alternating between generating EPR pairs and performing entanglement swapping over Bob, each swapping iteration sequentially opens up previously occupied qubits dedicated to EPR pairs from Alice. Multiple swapping iterations then allow for more than half of the qubits to be dedicated to EPR pairs from Alice, yielding a higher efficiency per satellite overpass.


Scaling the system further can be done by multiplexing many nodes in the satellite. Primarily, each node can be spectrally multiplexed either by having communication ions with slightly different resonance frequencies to emit spectrally separated photons, or by utilising non-linear processes such as frequency sum generation to convert the frequency of emitted photons with small to negligible losses \cite{Donohue2015,Fisher2021}. As the photon emission time is short in comparison to the qubit preparation \cite{Kinos2025}, temporal multiplexing also becomes a possibility by having several nodes read out photons at a small delay, filling the gap between the early-late photon pair emitted from each node.

Due to the long coherence time of the qubits in this scheme, entanglement can be set up in advance and only needs to be replaced once it has been used or decohered. An orbital band of satellites using this scheme would therefore not be limited by the time it takes a satellite to travel between the ground stations. Instead, each satellite would contribute to the distribution rate and would allow for continuous entanglement distribution.

\begin{figure}
\hspace{2ex}
\scalebox{0.98}{\input{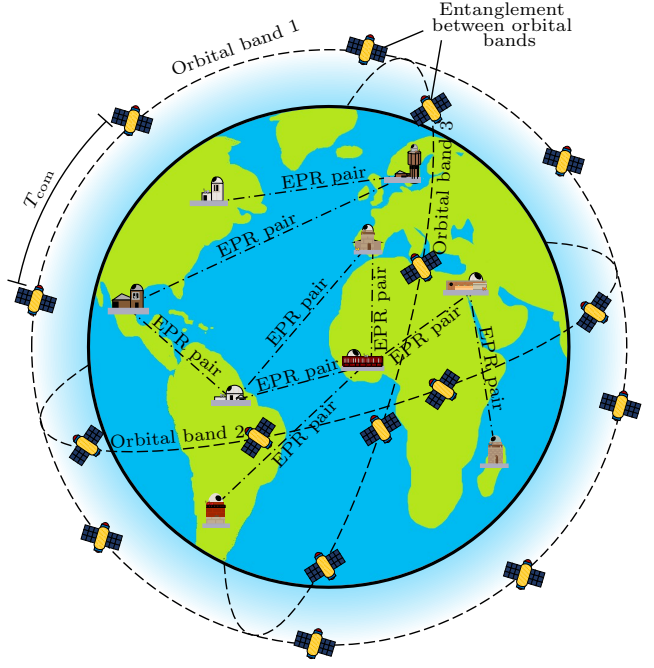}}

    \caption{Constellation of satellites allowing for global coverage without additional losses. }
    \label{fig:constellation}

\end{figure}
With an orbital band of satellites, each satellite could also entangle different ground stations with a common ground station, as illustrated in Figure \ref{fig:constellation}. Entanglement can then be set up between any ground station on demand by performing entanglement swapping with the qubits in the common ground station. This is only limited by the classical communication time between ground stations. Several orbital bands can then be connected either through a common ground station or by generating EPR pairs directly between satellites during a close pass, allowing for global coverage.

\begin{figure*}[t!]
\begin{tikzpicture}
    \node[anchor=south west,inner sep=0] (image) at (0,0) {\includegraphics[width=0.36\linewidth]{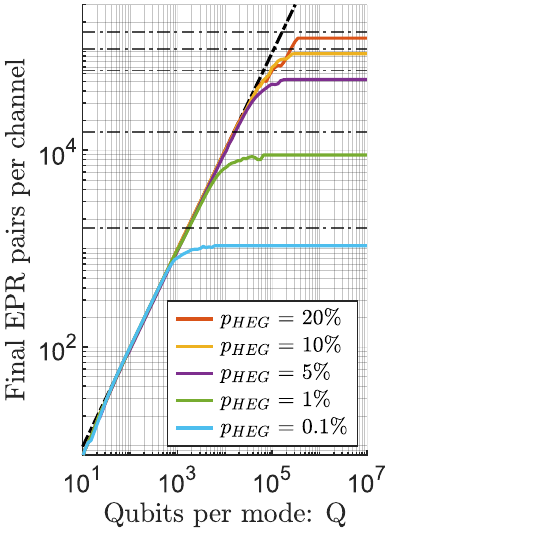}};
    \node[anchor=south west,inner sep=0] (image) at (0,0) {\hspace{30ex}\includegraphics[width=0.352\linewidth]{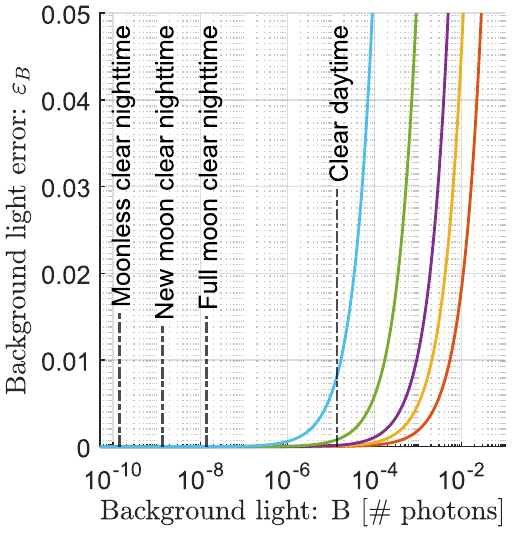}};
    \node[anchor=south west,inner sep=0] (image) at (0,0) {\hspace{72ex}\includegraphics[width=0.37\linewidth]{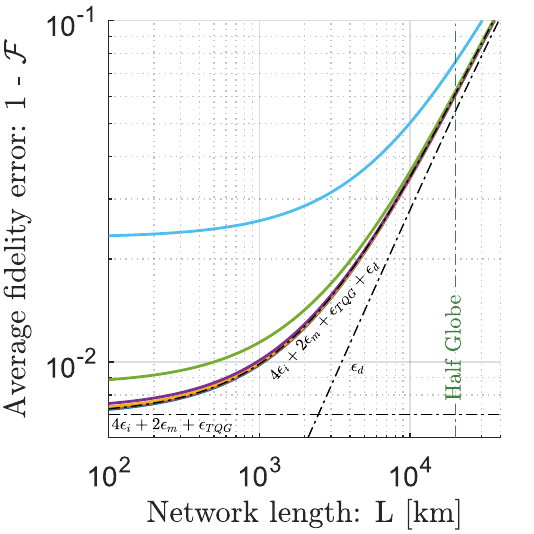}};
    \begin{scope}[x={(image.south east)},y={(image.north west)}]
        \node[black, font=\bfseries] at (0.019,1.02) {a)};
        \node[black, font=\bfseries] at (0.275,1.02) {b)};
        \node[black, font=\bfseries] at (0.65,1.02) {c)};

        \draw[black,<-] (0.94,0.82) -- (0.9,0.82);
        \node at (0.86,0.82) {\small{$\mathcal{F} \approx 94\%$}};
    \end{scope}
\end{tikzpicture}

\caption{Performance of our scheme with respect to different parameters. In all figures the line colors represent different HEG success probabilities, as defined in the legend of part a). $\eta_0 = 0.8$ is assumed.  
\textbf{a)} Simulated average number of EPR pairs distributed for a satellite with $Q$ qubits \textit{without multiplexing}. In the linear region, over $98\%$ of the qubits end up as EPR pairs, regardless of channel efficiency. Black horizontal lines denote the theoretical maximum of distributed EPR pairs in accordance with Equation \eqref{eq:knee}. It is assumed that $t_{\text{HEG}} = 20\ \unit{\mu s}$, $t_{\text{swap}} = 210\ \unit{\mu s}$ \cite{Kinos2025} and a satellite altitude of $250\ \unit{km}$, corresponding to a VLEO orbit. At this altitude, a trial efficiency of $p_{\text{HEG}} = 17\%$ could be achieved. A communication window of $T_{\text{com}} = 65\ \unit{s}$ is used, corresponding to the temporal full-width half maximum of $p_{\text{HEG}}$ in the given orbit. The simulation is described in detail in Methods \ref{appendix:Markov_Chain}-\ref{appendix:Time_Binning}. 
\textbf{b)} Fidelity error for a given average number of background light photons $B$ in an HEG trial and for different HEG trial success probabilities $p_{\text{HEG}}$. Black lines denote different typical lighting conditions.
\textbf{c)} Network length dependence of the final EPR pair error during clear daytime. A coherence time of $T_2 = 13\ \unit{h}$ and an error of $\epsilon_i=\epsilon_m=\epsilon_{TQG}=10^{-3}$ is assumed.}
\label{fig:simulation_and_fidelity}
\end{figure*}
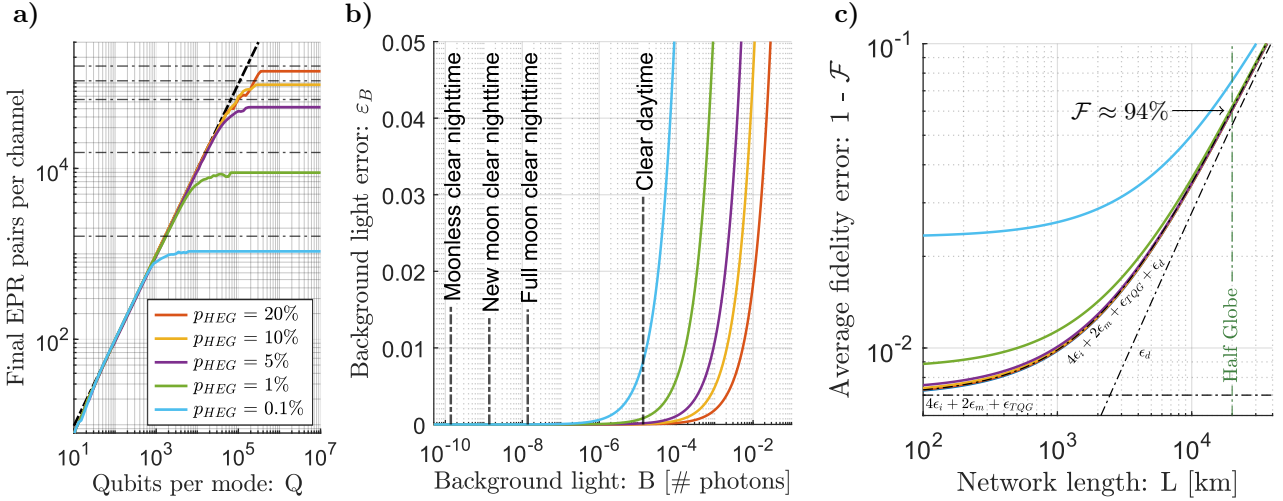

\subsection*{Entanglement distribution rates}

To estimate the maximal number of EPR pairs that can be generated, the following satellite overpass will be considered. A satellite empties the EPR pairs from a previous ground station by alternating between generating EPR pairs and performing swapping. Before finishing the overpass, new EPR pairs are generated with the current ground station, allowing for the scheme to be repeated at the next ground station. As seen in Figure \ref{fig:simulation_and_fidelity} a), close to all qubits can be dedicated to Alice with enough time to swap all states, yielding a final number of EPR pairs corresponding to up to $98\%$ of the total number of qubits used. If enough qubits are used without multiplexing, the efficiency starts to decrease, as there is no longer enough time to address all qubits before completing the overpass. For lower $p_{\text{HEG}}$, more time is needed for HEG trials to generate the same number of EPR pairs and the efficiency therefore decreases earlier for lower $p_{\text{HEG}}$. Above this limit, there is no need for additional qubits as the performance is now limited by the communication window and $p_{\text{HEG}}$, creating a plateau in performance. The plateau corresponds to the situation where half of the communication window is dedicated to generating EPR pairs with each of the ground stations and performing entanglement swapping once. The maximal number of EPR pairs that can be distributed is then given by:
\begin{align*}
    N = \frac{T_{\text{com}}}{\frac{2\cdot t_{\text{HEG}}}{p_{\text{HEG}}} + t_{\text{swap}}}
    \tag{2}
    \label{eq:knee}
\end{align*}
where $T_{\text{com}}$ is the duration of the communication window in one overpass, $t_{\text{HEG}}$ is the time to perform one HEG trial and $t_{\text{swap}}$ is the time needed to perform entanglement swapping. Here, $p_{\text{HEG}}$ is the average HEG trial efficiency during the satellite overpass.

By using a VLEO orbit of $250\ \unit{km}$, the optimised HEG trial efficiency is $p_{\text{HEG}} = 17\%$, if it is assumed that aperture of the transmitter (receiver) telescope is $0.8\ \unit{m}$ ($1.8\ \unit{m}$), the pointing error is $\theta_{p} = 1\ \unit{\mu rad}$, adaptive optics is used and the communication ion is neodymium with emission at $883\ \unit{nm}$.

\begin{table*}[t!]
\centering
\begin{tabular}{|c|c|c|c|c|}
    \hline
    & Rate & Fidelity & Second generation & Main problem \\
    & & & compatible & \\
    \hline
    Transportable qubits & High & High & Yes & Complexity \\
    \hline
    Transportable QM \cite{Gndoan2024,Wittig2017} & High & High & No & Inefficient photon storage \\
    \hline
    Satellite relay \cite{Goswami2023,Liu2026} & Low & Very low & No & Extreme losses \\
    \hline
    Balloon relay \cite{Liu2026} & Decent & Low & No & High losses \\
    \hline
    Satellite BSM with QM \cite{Gndoan2021, Liorni2021, Khatri2021} & Decent & Very high & No & Many satellites required \\
    \hline
    Ground BSM with QM \cite{Boone2015} & Decent & Very high & No & Many ground stations required \\ 
    \hline
\end{tabular}
\caption{Comparison of the main properties and drawbacks of different entanglement distribution schemes. QM denotes quantum memory and BSM denotes Bell state measurement. See the \textit{Rate comparison} section in the main text for more details.}
\label{tab:comparison}
\end{table*}

\subsection*{EPR pair fidelity}

To first order in each error type, the final EPR pair fidelity can be estimated as:
\begin{align*}
    \mathcal{F} = 1 - \qty[4\epsilon_i + 2\epsilon_m + \epsilon_{TQG} + \epsilon_d  + 2\epsilon_B]
    \tag{3}
    \label{eq:Final_EPR_fidelity}
\end{align*}
with $\epsilon_i$ being a qubit initialisation error occuring four times since one EPR pair is created at each ground station. $\epsilon_m$ is the probability of acquiring the wrong classical information upon measuring a quantum state in the entanglement swapping scheme and $\epsilon_{TQG}$ is the fidelity error of any two-qubit gate. $\epsilon_d$ is the decoherence error of the EPR pair carried by the satellite, as this is the only pair that exists for long enough to decohere significantly. To perform entanglement swapping, one two-qubit gate and two measurements are needed. $\epsilon_B$ is the background light induced error of the individual EPR pairs created at each ground station, as seen in Figure \ref{fig:simulation_and_fidelity} b). The error is given by the relative level of background light and the probability of succeeding an HEG trial. For a shorter network length, the background light conditions will be the same over both ground stations and the error will come in twice. However, for a long enough network length, the conditions will be different over each ground station and the error will be dominated by the ground station with the worst level of background light. The error then only comes in once. The full expression for the final EPR pair fidelity is derived in detail in Methods \ref{appendix:Error_Estimation}-\ref{appendix:Background_Light} and shown in Figure \ref{fig:simulation_and_fidelity} c).

\subsection*{Rate comparison}

The communication rate of satellite protocols can in general be broken down into two fundamental parts; firstly, how many EPR pairs can ultimately be set up during a satellite overpass over a ground station, and secondly, what is the loss transferring those EPR pairs around the globe. The limit of the first one is the same for all satellite-based protocols, and the second one we will argue is much better for a transportation-based protocol. A brief comparison of the different satellite-based schemes mentioned in this section can be found in Table \ref{tab:comparison}.

Generally, the maximal number of spectral modes possible is limited by the bandwidth allocated to an EPR pair $\Delta\nu_{\text{EPR}}$ and the total grid optical bandwidth $\Delta\nu_{\text{grid}}$. To ensure a photon overlap of at most $0.1\%$ for photons of Lorentzian shape, the minimal bandwidth required is $\Delta\nu_{\text{EPR}} \approx 5/\Delta t_{\text{ph}}$, for a photon with a $1/e$ decay time $\Delta t_{\text{ph}}$. Similarly, the temporal mode spacing needed is $\Delta t_{\text{EPR}} \approx 7\Delta t_{\text{ph}}$. The total number of EPR pairs physically possible to generate during an overpass is then given by:
\begin{align*}
    N_{\text{tot}} = \frac{T_{\text{com}}}{\Delta t_{\text{EPR}}}\cdot \frac{\Delta\nu_{\text{grid}}}{\Delta\nu_{\text{EPR}}}\cdot p_{\text{HEG}} \lesssim \frac{1}{35}T_{\text{com}}\Delta\nu_{\text{grid}}\cdot p_{\text{HEG}}
    \tag{3}
    \label{eq:channel_capacity}
\end{align*}
and is independent of the duration of the emitted photon. This maximum holds true for any scheme transmitting photons through the atmosphere with the given temporal and spectral mode separations.

Just as an example of future capacity, using a grid with similar bandwidth to the $1550\ \unit{nm}$ telecom ITU grid with $\Delta\nu_{\text{grid}}\approx 5\ \unit{THz}$, an orbital band with one satellite communicating during each $T_{\text{com}}$ would be capable of distribution rates at most on the scale of hundreds of GHz using a lossless system in accordance with Equation \eqref{eq:channel_capacity}. With a full orbital band of less than one hundred satellites, a continuous distribution rate on the scale of tens of GHz should therefore be possible in VLEO with $p_{\text{HEG}} = 17\%$.

For a qubit-based scheme, scaling the rate to tens of GHz would require approximately $10^{12}$ qubits per satellite and would of course pose a massive technical challenge to develop. However, only a few of the qubits need to be interconnected and many smaller processor nodes can be used in parallel to reach the desired number of qubits. Regardless, creating the processors needed for this type of scheme will be a complex task.

In the case of transportation-based schemes using multi-mode quantum memories without qubits \cite{Gndoan2024,Wittig2017}, entangled photon pairs can be stored in the ground station and the satellite with in principle identical losses to the qubit-based protocol. However, faulty photon pairs where one of the photons are lost along the channel will occupy most of the memory capacity as long as there are channel losses. Even if the faulty photons can be removed, the memory slots will be left empty instead, as all current proposals for long-lived quantum memories store photons in a \textit{first in, first out} manner. To fully fill a quantum memory in a transportation-based scheme, all successful photons would need to be retrieved, sorted in time, and added back to the memory multiple times. This is a highly inefficient process that would lower both the fidelity and rate of the scheme, which means that transportation-based schemes using memories may have to live with inefficient memory usage and instead have bigger memories.

In the case of satellites using direct relays, such as mirrors without memories \cite{Goswami2023}, each inter-satellite link would contribute to additional losses. The link distance needs to stay within two Rayleigh lengths in order to keep losses from beam diffraction low, meaning that any link distances should be kept below a thousand kilometers or telescope mirrors becomes several meters in diameter which appears unfeasible. Tens of satellites will therefore be needed to reach global distances. At high altitudes, channel losses between the ground and the satellite will be significant and the highest achievable rates should therefore be found in VLEO. At lower altitudes, a higher number of satellites will be needed in order to keep $p_{\text{HEG}}$ from varying with the transmission angle during the overpass. Since a photon will have to survive the entire chain at once, the channel losses will be significant, pushing the signal closer to the level of background light. Both the rate and fidelity in this type of scheme should therefore be orders of magnitude lower than in the transportable case. This can be somewhat negated by using balloons as relays at even lower altitudes \cite{Liu2026}. However, the losses are still higher than transportable schemes and balloons will have a comparably lower rate and fidelity.

With access to highly efficient quantum memories, one can instead use the satellites as quantum repeaters, directly connecting photons emitted from each link with Bell state measurements, similar to fiber-based protocols on the ground \cite{Gndoan2021, Liorni2021, Khatri2021}. If the link loss is similar or lower than the satellite to ground loss, it should be possible to distribute EPR pairs at the maximal rate limited by the two downlink channels. However, entanglement swapping using Bell state measurements are limited to $50\%$ success probability. Thus, if $M$ links are used, $\mathrm{log}_2\qty(M)$ Bell state measurements are needed in the optimal case. This will introduce an additional loss of $\qty(1/2)^{\mathrm{log}_2\qty(M)} = M^{-1}$, leading to lower but still decent rates. However, since photons are only stored in quantum memories briefly in this scheme, decoherence is negligible and the only fidelity error stems from background light. The fidelity of this type of scheme will therefore be higher than for the transportable case.

Another version of this protocol is that the satellites send their photons to ground stations for the Bell state measurement, instead of other satellites \cite{Boone2015}. The scaling would be identical, although with slightly increased downlink losses due to having to transmit photons to the ground at an angle other than the zenith at all times. A global distance network would in addition require a multitude of ground stations in a line along the orbital path, around the earth. Geological features such as oceans would therefore highly impact the technical difficulty of implementing such a scheme.
\section*{Discussion}

Considering a realistic system based on REIs, up to one hundred qubits per communication ion can theoretically be expected for a laser with $100\ \unit{GHz}$ tuning range \cite{Kinos2022}. This is enough qubits to run the scheme, and a larger memory size can be made from many rare-earth nodes working in parallel. Similarly, entanglement swapping of qubits connected to one communication ion can be done in parallel to the HEG trials on other communication ions, yielding an effective swapping time close to zero.

To sustain a high distribution rate in an orbital band, enough satellites for $p_{\text{HEG}}$ not to vary substantially is needed. Using a one minute communication window, less than one hundred satellites are needed to fill the orbital band and $p_{\text{HEG}}$ varies less than one full-width half-maximum during the overpass in accordance with Method \ref{appendix:Channel_Losses}. However, it is possible to start with a single satellite and then gradually increase the number of satellites with increasing demand, ultimately reaching a rate of tens of GHz per orbital band, assuming that each satellite performs at the theoretical maximum.
For near term systems with fewer qubits, each satellite would be able to deliver a bulk of EPR pairs corresponding to $98\%$ of the qubit capacity $Q$. For a full orbital band, the rate would be approximately $Q/T_{\text{com}}$.

Achievable HEG trial efficiencies are sufficiently high that the fidelity is negligibly impacted by background light both during day- and night-time operation, as seen in Figure \ref{fig:simulation_and_fidelity} b). The fidelity for the final EPR pairs when distributed globally is then limited to approximately $\mathcal{F} = 93.9\%$ and limited to $\mathcal{F} = 99.3\%$ at shorter distance where decoherence is negligible. This assumes a coherence time of $T_2 = 13\ \unit{h}$ and errors of size $10^{-3}$ \cite{Kinos2021_2}.
The errors are well below the limit to perform QKD schemes such as BB84 \cite{Shor2000} and if a higher fidelity is desired, entanglement purification can be performed at the cost of sacrificing half of the EPR pairs per purification round \cite{Deutsch1996, Dr1999}. This holds true both for initial and final EPR pairs, meaning that a higher fidelity requirement can be achieved with a reduced distribution rate.

It is noted that the fidelity of the initial EPR pairs between the satellite and one of the ground stations, $\mathcal{F} = 99.6\%$, is significantly higher than the fidelity of the final EPR pair since the initial pairs are only limited by initialisation errors and the background light induced error. This is high enough to allow for certain error correction protocols to be implemented directly between ground stations \cite{Raussendorf2007}, thus making our scheme the first to be compatible with second (and potentially third) generation communication systems \cite{Muralidharan2016}. Using error correction can reduce the impact of decoherence errors on the EPR pairs traveling from Alice, thus improving the final EPR pair fidelity.


In conclusion, we have shown a new scheme for distributing entanglement globally with the use of qubits. We have shown that the scheme can distribute close to the theoretical number of EPR pairs and that it in theory will outperform other satellite-based schemes. The high channel efficiencies in VLEO makes daytime operation feasible, with a fidelity high enough to perform entanglement purification or even certain error correction protocols. Lastly, combining the scheme with a satellite constellation and inter-satellite communication would in turn allow for entanglement distribution at global distances with global coverage.

\section*{Acknowledgements}

This work was supported by the Knut and Alice Wallenberg Foundation through the Wallenberg Centre of Quantum Technology (WACQT).

\printbibliography

\section*{Methods}
\renewcommand{\thesubsection}{\Alph{subsection}}
\subsection{Channel efficiency}
\phantomsection
\label{appendix:Channel_Losses}

The probability of detecting an emitted photon for a given path is determined by the channel-independent efficiency $\eta_0$ and the channel efficiency $\eta_c$. $\eta_0$ accounts for the efficiency of emitting a photon, coupling it into the detection system, and the detector's efficiency. This efficiency typically ranges between $5-80\%$ \cite{Kinos2025}. $\eta_c$, accounts for photon losses during propagation.

For photons emitted within the ground station, the detection efficiency is only limited by $\eta_0$. However, for a photon emitted from the satellite, $\eta_c$ must be included, so that the overall probability of detecting one photon from each station is $\eta_0^2\eta_c$.

In this section, the channel efficiency $\eta_c$ was investigated by studying the main effect that cause photons to be lost as they travel from the satellite to the ground station:
\begin{align*}
    \eta_c = \eta_t \eta_r \eta_{\mathrm{at}} \eta_{\mathrm{SMF}}
    \tag{A.1}
\end{align*}
where $\eta_t$ is the transmitter efficiency, $\eta_r$ is the receiver efficiency, $\eta_{\mathrm{at}}$ is the atmospheric transmission efficiency, and $\eta_{\mathrm{SMF}}$ is the coupling efficiency into a single mode fiber (SMF).


\subsubsection{Transmitter and receiver efficiencies}

Since the transmitted beam has a Gaussian distribution and both the transmitter and receiver telescope have finite apertures, a small portion of the beam power must be truncated, resulting in losses. In addition to this, the beam size changes during its propagation from the satellite to the ground station, which can further reduce the optical power collected by the receiver. 

For a transmitted beam with a radius of $\omega_t$ and a transmitter aperture radius of $r_t$, the transmitter efficiency $\eta_t$ can be determined as follow:
\begin{align*}
    \eta_{t} = 1-e^{-2r_t^{2}/\omega_{t}^{2}}
    \label{geo_eta}
    \tag{A.2}
\end{align*}

Similarly, for an optical receiver with aperture radius $r_r$, the receiver efficiency $\eta_r$ is given by:
\begin{align*}
    \eta_{r} = 1-e^{-2r_r^{2}/\omega_{r}^{2}}
    \label{rec_eta}
    \tag{A.3}
\end{align*}

where $\omega_r$ is the effective beam radius at the receiver, which can be determined from the effects that causes the beam size to change as it travels through free-space. This could either be due to beam diffraction, atmospheric turbulence, or beam wandering due to pointing error in the transmitter.

First the transmitted beam is focused at a distance $R_0$ from the transmitter, which gives a beam waist radius $\omega_0$ of:
 \begin{align*}
 \begin{aligned}
     \omega_{0} = \sqrt{\frac{\omega_t^2 + \sqrt{\omega_t^4 -4\left(\frac{R_0 \lambda}{\pi}\right)^2}}{2}}
 \end{aligned}
     \label{divergancer}
     \tag{A.4}
 \end{align*}
After the focus the beam will spread during propagation due to diffraction. 
The beam size after a distance $R'$ can be calculated as follow:
 \begin{align*}
 \begin{aligned}
     \omega_{d} = \omega_{0}\sqrt{1 + \left(\frac{R'-R_0}{z_R}\right) ^2}
 \end{aligned}
     \label{divergancer}
     \tag{A.5}
 \end{align*}

The distance from the satellite to the ground station ($R'$) changes as the satellite moves in its orbit, due to the change in the zenith angle between the satellite and the ground station. The effective distance can be approximately found with respect to the zenith angle $\theta_{z}$ as follow:
  \begin{align*}
     R' = R_{\mathrm{sat}}\ \mathrm{sec}(\theta_z)
     \label{distance}
     \tag{A.6}
 \end{align*}
 where $R_{\mathrm{sat}}$ is the satellite's altitude.
 
The beam traveling through the atmospheric turbulence will experience further broadening due to random variations in temperature and pressure in the atmosphere leading to fluctuations in the refractive index. This turbulence can be thought of as lenses that can cause two effects based on their size compared to the beam. If the turbulence is larger than the beam itself, that causes beam wandering, whereas, if the beam size is larger, that causes beam broadening. For a satellite-to-ground link, the beam travels through the atmosphere only in the final path of its propagation where its size large enough to make the beam wandering effect negligible \cite{bonato2009feasibility}. The resulting beam radius can be calculated as follow:  
    \begin{align*}
\begin{aligned}
    \omega_{\mathrm{tur}}= \sqrt{\omega_{d}^{2} + 2\left(\frac{4R'}{k r_{0}}\right)^{2}} \\
\end{aligned}
\label{broadening}
\tag{A.7}
\end{align*}
where $\omega_{d}$ is the beam size due to diffraction shown in Eq. \eqref{divergancer}, $k$ is the wave number, and $r_{0}$ is the Fried parameter and can be found for a downlink as follow \cite{maharjan2022atmospheric, kaushal2017experimental}:
\begin{align*}
\begin{aligned}
    r_{0}= 1.67 \left[ \mathrm{sec(\theta_{z})} k^{2} \int_{0}^{R'} C_{n}^2(h)dh\right]^{-\frac{3}{5}} \\ 
\end{aligned}
\label{fried parameter}
\tag{A.8}
\end{align*}
here, $C_{n}^2$ is the refractive index structure parameter, which quantifies the strength of atmospheric turbulence. The value of $C_{n}^2$ scales with the altitude, and is larger closer to the ground due to a larger gradient of temperature, and lower closer to the upper atmosphere \cite{altowij2010effect}. Several models have been proposed to describe $C_{n}^2(h)$ with the most popular one being the following Hufnagel-Valley Model \cite{valley1980isoplanatic}.
\begin{align*}
\begin{aligned}
    C_{n}^2(h) &= 0.00594\left(\frac{v_{\mathrm{wind}}}{27}\right)^{2}\left(10^{-5}h\right)^{10} e^{-\frac{h}{1000}} \\ 
    & + 2.7\times 10^{-16} e^{-\frac{h}{1500}} + A_{0}e^{-\frac{h}{100}} \\
\end{aligned}
\label{refractive index structure parameter}
\tag{A.9}
\end{align*}
where $h$ is the altitude in $\mathrm{m}$, $v_{\mathrm{wind}}$ is the wind speed in $\mathrm{m/s}$, and $A_0$ is the nominal value of $C_{n}^2(0)$ in $\mathrm{m^{-2/3}}$.
For weak turbulence conditions, $C_{n}^2$ is less than $10^{-17}$ $\mathrm{m^{-2/3}}$. For moderate turbulence condition $C_{n}^2$ range from $10^{-17} - 10^{-13}\ \mathrm{m^{-2/3}} $, and for strong turbulence condition  $C_{n}^2$ is higher than $10^{-13}\ \mathrm{m^{-2/3}} $ \cite{andrews2023laser}. 
\\

Finally, free-space optical (FSO) communication requires a line of sight between the fast-moving satellite and the ground station. Any misalignment between the transmitter and receiver introduces additional losses. This misalignment can be caused by satellite jitter resulting from altitude variation and mechanical vibrations during its orbit motion, which causes the beam center to deviate from the nominal line connecting the transmitter and receiver. This can be modeled as a further bream broadening such that the effective beam radius can be found as follow:
\begin{equation}
    \omega_{r} = \sqrt{\omega_{\mathrm{tur}}^{2} + 4\sigma_{P}^{2}}
    \tag{A.10}
\end{equation}
Here, $\sigma_{p} \simeq R'\theta_{p}$ is the pointing error standard deviation and is characterized by the pointing angle $\theta_{p}$ \cite{dequal2021feasibility}. The receiver efficiency $\eta_r$ can then be calculated using Eq. \eqref{rec_eta}.

To minimize signal loss caused by pointing errors, a high-precision and high-speed tracking system is required. Most FSO satellites employ an Acquisition, Tracking, and Pointing (APT) system to maintain accurate beam alignment. This system enables a pointing error angle $\theta_p$ on the order of few microradians or even sub-microradians accuracy.


\subsubsection{Atmospheric transmission efficiency}
Small particles and gas molecules in the atmosphere introduce additional losses. Gas molecules can absorb photons whose wavelengths correspond to their characteristic absorption lines or Rayleigh scattering, whose strength depend on the photon wavelength. Additionally, small particles, such as water droplets, dust, and aerosols, scatter photons through Mie scattering, which also depends on the particle size relative to the wavelength.

The type and concentration of molecules and particles presents in the atmosphere also effect the atmospheric absorption, making it dependent on the location, altitude, time of day, and the weather conditions. These molecules and particles mostly exists in the last portion of the atmosphere. Therefore, losses due to atmospheric absorption is constant for large altitudes. 

 Since atmospheric losses depends on the amount of gas molecules and  small particle that the beam encounters in its path, it is a function of the effective distance and can be calculated as a function of the zenith angel as follow \cite{dequal2021feasibility}, where $\eta_{at(z)}$ is the transmission efficiency at the zenith: 
 \begin{align*}
     \eta_{\mathrm{at}}(\theta_{z}) = {\eta_{\mathrm{at}}}^{\mathrm{sec(\theta_{z})}}
     \label{at_eta}
     \tag{A.11}
 \end{align*}
Different software were developed to determine the atmospheric transmission for a certain wavelength, including MODTRAN \cite{modtran5_ontar2010}. By using the MODTRAN software to estimate $\eta_{\mathrm{at}}$ for a typical rural location, with a clear-sky conditions, it is found that for a communication ion of Nd (doped into a \(\text{Y}_{2}\text{SiO}_{5}\) crystal) with an emission at around $883\,\text{nm}$, the transmission efficiency is $\eta_{\mathrm{at}} \approx 83\%$.

 \subsubsection{Coupling efficiency into a single mode fiber}
 Before detection, the photon emitted within the ground station and the photon arriving from the satellite must be indistinguishable for the entanglement generation to succeed. This also requires a high spatial mode overlap between the photons. To achieve this, adaptive optics are used to correct the wavefront distortion in the beam received from the satellite. 
 
 For a downlink channel, since the beam travels almost entirely through vacuum, it experiences primarily broadening due to diffraction during most of its propagation path. The beam encounters atmospheric turbulence only in the final stage before reaching the receiver, meaning that turbulence-induced aberrations have a comparatively weak effect on the wavefront prior to entering the telescope which can be corrected for using adaptive optics \cite{bonato2009feasibility}. 
 
 The beams are coupled into a single mode fiber (SMF) to ensures spatial mode overlap between the photons. The coupling efficiency into a SMF $\eta_{\mathrm{SMF}}$ is determined by the overlap integral between the received Gaussian field and the fundamental mode of fiber. For an ideal Gaussian beam with perfect mode matching, the theoretical coupling efficiency can approach $\eta_{\mathrm{SMF}}= 99.8\%$ for a gaussian mode \cite{miller2012optical}. In practice, With appropriate mode-matching optics, coupling efficiencies of up to $\eta_{\mathrm{SMF}}\approx 90\%$ has been demonstrated \cite{Knothe2021SMF}.
\subsubsection{Expected system efficiency:}

This section outlines feasible parameters to give quantitative values for the system efficiencies. The parameters are selected to ensure a feasible and physically consistent model. The baseline configuration relies on existing or developing technology. For instance, the satellite is assumed to be in the VLEO with an altitude of $250\, \mathrm{km}$, which ensures less losses due to beam divergence. 

The specifications for the optical beam, such as wavelength and beam waist size, can greatly influence $\eta_c$. If instead of using a Nd communication ion with an emission at around $883\,\text{nm}$, Er with an emission around $1536\, \text{nm}$ is used, the transmission efficiency $\eta_{\mathrm{at}}$ improves from  $83\%$ to $93\%$. However, Er has a higher wavelength, resulting in a higher divergence which reduces $\eta_{g}$ and the overall $\eta_c$.

Another important parameter to take into consideration is the beam waist radius $\omega_0$, as a smaller beam radius at the receiver results in a higher fraction of optical power being coupled into the receiver aperture, leading to an increase in $\eta_{g}$, therefore, 
\if 0\color{red} If the beam waist is too small, the beam diverges rapidly, resulting in a larger beam size at the ground station. As the beam waist size increases, the beam divergence decrease, and received beam waist becomes smaller. This trend will continue until an optimal point is reached, for which the Rayleigh range becomes comparable with the propagation distance $L=z_{R} =\frac{\pi w_{0}^{2}}{\lambda}$. Beyond this pint, further increasing the initial beam waist will cause the received beam size to increase again.
The relationship between the beam size at the ground station and the channel efficiency is non-trivial, as the geometrical efficiency and the pointing error efficiency exhibit opposite trends. A small beam waist results in a higher fraction of optical power being coupled into the receiver aperture, leadineg to an increase in $\eta_{g}$. However, that will make the system more sensitive to misalignment and thus reduce $\eta_{p}$. \fi
$\omega_0$ must be optimized. For instance, when the beam waist is located at the center between the satellite and the ground station, the maximum $\eta_{c}$ occurs when the beam waist $\omega_0$ is chosen such that the Rayleigh length is comparable to half the link distance, i.e., ($w_0=\sqrt{\frac{\lambda R}{2\pi}}$). The location for the ground-station also plays a important role in reducing beam broadening due to atmospheric turbulence. The parameters for the Hufnagel-Valley model in \eqref{refractive index structure parameter} can be found to be $A_0 = 1.7$$ \times 10^{-14}$ and $v_{\mathrm{wind}} = 21\ \mathrm{m/s}$ for a ground-station close to sea level, whereas, placing the ground-station at a more favorable location can help reduce $A_0$ by at least an order of magnitude.

Lastly, system parameters such as the aperture diameter for both the satellite and the ground-station telescopes play an important role for $\eta_c$. Small satellites systems, such as  MIT-PorTeL \cite{riesing2018portable} and TeraNet-3 \cite{walsh2025teranet}, have relatively small ground-station telescope apertures of around $28-40 \ \mathrm{cm}$, whereas, systems with larger optical ground stations, such as the Micius \cite{lu2022micius,yin2017satellite} and OPALS\cite{abrahamson2015achieving}, employ telescope apertures of approximately $1-1.8\ \mathrm{m}$. Similarly, the telescope size onboard the satellite can be as small as $18-30 \ \mathrm{cm}$ for systems such as \cite{lu2022micius,yin2017satellite}, while others such as Pléiades Neo \cite{jerome2019shaping} and worldview-3 \cite{worldview3_eoportal} may use apertures exceeding $1\ \mathrm{m}$. These variations can significantly influence $\eta_c$, from around $0.5\%$ up to $50\%$.

Taking all of the above into account, we assume a satellite telescope aperture radius of $r_t=0.4\ \mathrm{m}$, resulting in a transmitter efficiency $\eta_{t}\approx 98.9\%$. A ground-station aperture radius of $r_r=0.9\ \mathrm{m}$, and the ground station is assumed to be located in a region with low-to-moderate atmospheric turbulence, characterized by $A_0=1.7\times10^{-15}$. We further assume a pointing error angle of $\theta_p=1\ \mathrm{\mu rad}$ \cite{dequal2021feasibility,yin2017satellite}, leading to a receiver efficiency of $\eta_{r}= 70\%$. For a Nd communication ion, we obtain an atmospheric transmission efficiency of  $\eta_{\mathrm{at}} = 83\%$. A SMF coupling efficiency of $\eta_{\mathrm{SMF}} =90\%$ was assumed. Combining these contributions yields an overall channel efficiency of $\eta_c \approx 52\%$. When assuming a channel-independent efficient of $\eta_0 \approx80\%$, this corresponds to a detection probability of $ \approx 41\%$ for photons emitted from the satellite towards the ground station, and $\approx 80\%$ for photons emitted within the ground station. If the effect of background light on the rate is neglected, the HEG trial success probability becomes $p_{\mathrm{HEG}}=\frac{1}{2}\eta_0^2\eta_c \approx 17\% $.
\\ 
\subsection{Simulating an overpass using Markov chains}
\label{appendix:Markov_Chain}


To estimate the average number of EPR pairs that can be generated by this scheme during an overpass, a Markov chain is utilised. The chain determines the probability of generating a number of EPR pairs given a number of already occupied qubits in the satellite node and is generated with the conditional probabilities $p\qty(s_i|s_{i-1})$ to have generated a total of $s_i$ EPR pairs in HEG trial round $i$ given $s_{i-1}$ pairs in the previous round. This is given by the binomial distribution:
\begin{align*}
    p\qty(s_i|s_{i-1}) = \begin{pmatrix}
        M_i \\
        k_i
    \end{pmatrix} p_{\text{HEG}}^{k_i}\qty(1- p_{\text{HEG}})^{M_i-k_i}
    \label{eq:Appendix_B_ConditionalProbability}
    \tag{B.1}
\end{align*}
where the number of HEG trials $M_i = Q - s_{i-1}$ in each round $i$ is maximized using all qubits $Q$ in the satellite that are currently not occupied by an EPR pair and $k_i = s_i - s_{i-1}$ is the number of successful HEG trials in the current round. On average, $\expval{k_i} = \expval{M_i}\cdot p_{\text{HEG}}$ pairs will be generated in each round for a constant $p_{\text{HEG}}$, resulting in:
\begin{align*}
    \expval{s_N} = \sum_{i=1}^{N}\expval{k_i} = Q\qty[1 - \qty(1-p_{\text{HEG}})^N]
    \label{eq:Appendix_B_ExpectedNumberStates}
    \tag{B.2}
\end{align*}
EPR pairs being generated in $N$ rounds of HEG trials. Consequently, the number of rounds $N$ needed to generate EPR pairs with a certain fraction $s_N/Q$ of the qubits is independent of the size of $Q$. Solving the problem with the Markov chain yields an identical mean value, the only difference being the resolution of the state space of possible values $s\in \qty{0,1,...,Q}$. By choosing a sufficiently high resolution, the computational time of the Markov chain can be reduced while yielding the same result. Through testing of the Markov chain, it can be seen that the selection of $Q = 100$ is sufficient and is used for $Q > 100$ to solve the problem before re-normalizing to the actual number of qubits. For $Q\leq 100$, the actual number of qubits is used to be able to see the effects of having few available qubits. For a $p_{\text{HEG}}$ varying between rounds, the lower resolution Markov chain can still be used given that the correct $p_{\text{HEG}}$ for each event can be calculated.

To simulate the overpass over Alice, the satellite is assumed to contain no EPR pairs and takes an input state $s_0 = 0$. It then iterates $N_A$ rounds through the Markov chain, yielding a distribution $p\qty(s_A)$ of the number of EPR pairs $s_A$ generated between Alice and the satellite. To simulate the overpass over Bob, $s_A$ is used as input state for the chain and is iterated through $N_B$ rounds, which allows us to find the conditional probabilities $p\qty(s_B|s_A)$ for $s_B$ states being generated between Bob and the satellite. The joint state distribution of $s_A$ and $s_B$ is then given by:
\begin{align*}
    p\qty(s_A,s_B) = p\qty(s_A)p\qty(s_B|s_A)
    \tag{B.3}
    \label{eq:joint_state_distribution}
\end{align*}
Entanglement swapping is then performed, requiring one EPR pair from $s_A$ and one from $s_B$. The maximum number of EPR pairs that can be swapped is $s_{\text{swap}} = \text{min}\qty(s_A,s_B)$ with the distribution:
\begin{align*}
    p\qty(s_{\text{swap}}) = \hspace{-2.5ex}\sum_{s_A=s_\text{swap}}^{Q-s_\text{swap}}\hspace{-2ex}p\qty(s_A,s_\text{swap}) + \hspace{-4ex}\sum_{s_B=s_\text{swap}+1}^{Q-s_\text{swap}}\hspace{-3ex}p\qty(s_\text{swap},s_B)
    \tag{B.4}
    \label{eq:probability_to_swap}
\end{align*}

The distribution of swapped EPR pairs $p\qty(s_{\text{swap}})$ is then saved and the remaining distributions are updated in accordance with: $s_A \rightarrow s_A - s_{\text{swap}}$ and $s_B \rightarrow s_B - s_{\text{swap}}$. If $s_A \neq 0$, another set of $N_B$ rounds can be performed to refill $s_B$ with the new updated $M_1 = Q - s_A$. Another set of entanglement swapping is then performed, adding the new swapped pairs from the latest swap to the probability distribution $p\qty(s_{\text{swap}})$ and once again updating $s_A$ and $s_B$. This is repeated $N_{\text{swap}}$ times, yielding a final distribution $p\qty(s_{\text{swap}})$. The optimal selection of $N_A$, $N_B$ and $N_{\text{swap}}$ is then found by maximising the average of $p\qty(s_{\text{swap}})$ on a discrete three-dimensional grid.
\subsection{Time-dependent HEG trial success rate}
\label{appendix:Time_Binning}

Due to the zenith angle $\theta_z$ varying as the satellite is performing its' overpass over a ground station, $p_{\text{HEG}}$ becomes time-dependent and each HEG trial needs to be time-binned. Going forward, it is assumed that all HEG trials occur fast enough for $p_{\text{HEG}}$ to be constant during a round of trials and that the zenith angle for that round can be approximated by the angle at the center of the time bin. To determine the time-dependent loss $p_{\text{HEG}}\qty(t) = p_{\text{HEG}}\qty(\theta_{z}\qty(t))$, it is assumed that the satellite travels in a circular orbit at distance $R_{sat}$ from the earth (with radius $R_E$) at a velocity $v$. The zenith angle is then defined as:
\begin{align*}
\begin{cases}
    \theta_{z}\qty(t) &= \text{arctan}\qty(\frac{\text{sin}\qty(\theta(t))}{\text{cos}\qty(\theta(t)) - R_E/(R_{\text{sat}}+R_E)}) \\
    \theta\qty(t) &= \frac{v}{R_E+R_{\text{sat}}}\cdot t
\end{cases}
\label{eq:Appendix_C_Angles}
\tag{C.1}
\end{align*}
where $\theta\qty(t)$ is the orbital angle formed from the center of the earth to the satellite position compared to being straight above the ground station, which occurs at $t=0$.

The total time $T_{\text{com}}$ that is needed for generating EPR pairs and performing entanglement swapping is heavily dependent on $p_{\text{HEG}}$ and the current number of qubits occupied by EPR pairs. It is therefore difficult to estimate how much time is needed without some sort of backpropagation algorithm capable of re-evaluating the time needed based on simulation results. However, the worst case scenario can be estimated as:
\begin{align*}
    T_{\text{com}} &= \qty(N_A+N_B\cdot N_{\text{swap}})\qty[Q\cdot t_{\text{HEG}} + t_{\text{rt}}\qty(\theta_{z}^{\text{max}})] \\
    &+ N_{\text{swap}}\cdot Q\cdot t_{\text{swap}}
    \label{eq:Appendix_C_T_Com}
    \tag{C.2}
\end{align*}
assuming that $t_{\text{HEG}}$ is the duration of an HEG trial, $t_{\text{swap}}$ is the duration needed for entanglement swapping and the roundtrip time for communication between the satellite and the ground station is $t_{\text{rt}}\qty(\theta_{z}) = \frac{2R_{\text{sat}}}{c} \text{sec}\qty(\theta_{z})$. The time-binning is then limited to the range $\qty|t|\leq T_{\text{com}}/2$, with maximum zenith angle $\theta_{z}^{\text{max}} = \theta_{z}\qty(T_{\text{com}}/2)$.

To simplify the simulation, we dedicate enough time for the worst case scenario. In reality, this will reduce the time that could have been used for more rounds of HEG trials, although the effect is negligible as long as the entire communication window is not being used. The simulation optimum will therefore deviate slightly from the theoretical maximum at large $Q$, predominantly in the cases of low $p_{\text{HEG}}$. 

The steady state solution for an overpass consists of a satellite first acting as a receiver, performing alternating $N_B$ rounds of HEG trials and one round of swapping a total of $N_{\text{swap}}$ times. It then refills with new states by performing $N_A$ rounds of HEG trials before leaving for the next ground station. By iterating through the events in that order, the center of each time bin is found, giving the values for $p_{\text{HEG}}$.
\subsection{Error estimation}
\label{appendix:Error_Estimation}

To estimate the fidelity of the final EPR pairs, an error analysis is made based on the semi-classical approach in Ref. \cite{Kinos2025} using density states $\rho = \ket{\psi}\bra{\psi}$ in the EPR state basis $\ket{\psi}\in\qty{\ket{\Phi^+}, \ket{\Psi^-}, \ket{\Psi^+}, \ket{\Phi^-}}$:
\begin{align*}
    \begin{cases}
        \ket{\Phi^+} = \frac{1}{\sqrt{2}}\qty(\ket{00} + \ket{11}) \\
        \ket{\Psi^-} = \frac{1}{\sqrt{2}}\qty(\ket{01} - \ket{10}) \\
        \ket{\Psi^+} = \frac{1}{\sqrt{2}}\qty(\ket{01} + \ket{10}) \\
        \ket{\Phi^-} = \frac{1}{\sqrt{2}}\qty(\ket{00} - \ket{11})
    \end{cases}
    \tag{D.1}
    \label{eq:Bell_states}
\end{align*}
Let's consider two EPR pairs $\ket{\psi_1}$ and $\ket{\psi_2}$ shared between the satellite and the two ground stations Alice and Bob, respectively. The density matrix $\rho_j \rightarrow \text{diag}\qty{A_j,B_j,C_j,D_j}$ will then have diagonal entries corresponding to the probabilities of acquiring any of the four EPR states. In the ideal case $C_j = 1,\ \forall\ j\in\qty{1,2}$, entanglement swapping would perfectly create a $\ket{\Psi^+}$ state. However, any amplitude in another state will result in errors accumulating in the scheme.

Upon EPR pair creation, a phase flip can occur due to an initialisation error $\epsilon_i$ on either of the two qubits, yielding the state:
\begin{align*}
    \rho_j &\rightarrow \qty[1-2\epsilon_i\qty(1-\epsilon_i)]\rho_j + 2\epsilon_i\qty(1-\epsilon_i)Z\qty(\rho_j)
    \tag{D.2}
    \label{Eq:Init_Error}
\end{align*}
with $Z\qty(\rho_j) \rightarrow \text{diag}\qty{D_j,C_j,B_j,A_j}$ being a phase flip operation. In addition, any qubit idling time will let the EPR pair decohere in accordance with:
\begin{align*}
    \rho_j &\rightarrow \qty(1-\epsilon_{d})\rho_j +\epsilon_{d}Z\qty(\rho_j)
    \tag{D.3}
    \label{Eq:Decoherence_Error}
\end{align*}
where $\epsilon_d = \qty(1-e^{-2t_j/T_2})/2$ is the decoherence error of the two qubits in an EPR pair with wait time $t_j$ and coherence time $T_2$. Taking both into account, each created EPR pair can be estimated to be in state:
\begin{align*}
    \rho_{j}^{in} &= \rho_j[1 - 2\epsilon_i\qty(1-\epsilon_i) - \epsilon_d + 2\epsilon_i\epsilon_d\qty(1-\epsilon_i)] \\
    &+ Z\qty(\rho_j)[2\epsilon_i\qty(1-\epsilon_i) + \epsilon_d - 2\epsilon_i\epsilon_d\qty(1-\epsilon_i)]
    \tag{D.4}
    \label{eq:Init_state}
\end{align*}
When performing entanglement swapping, there are two input states: $\rho_1^{in}$ from Alice with $t_1/T_2\approx L_o/vT_2$ and one from Bob $\rho_2^{in}$ with $t_2/T_2\approx 0$. Here $L_o$ is the distance between the two ground stations along the orbital path of the satellite and $v$ is the velocity of the satellite. Time variations between the different HEG trials are assumed to be negligible in comparison to the coherence time.


Similar to the error analysis made in Ref. \cite{Kinos2025}, entanglement swapping will generate two types of errors. The first one being a two-qubit gate error at a rate $\epsilon_{TQG}$ in the CNOT gate operation. This will in turn have a $\frac{\epsilon_{TQG}}{3}$ probability of flipping a Bell state into any of the other three Bell states and the average state is given by:
\begin{align*}
    \rho_{j}^{TQG} = \qty(1-\frac{4\epsilon_{TQG}}{3})\rho_{j}^{in} + \frac{\epsilon_{TQG}}{3}\mathrm{I}
    \tag{D.5}
    \label{Eq:TQG_Error}
\end{align*}
The second error is generated when performing a measurement on the two satellite qubits, where there is a probability $\epsilon_m$ that a classical bit flip error will occur in the measurement. As a result, $\rho_1$ will have a probability of acquiring a phase error from the controlled-Z gate and $\rho_2$ will have a probability of acquiring a bit flip error from the controlled-X gate. If $\epsilon_m$ is identical for both paths, these errors are equivalent to $\rho_2$ having the probability of acquiring a bit flip, phase flip or both. We therefore define the state:
\begin{align*}
    \rho_2^{m} &= \qty(1-\epsilon_m)^2\rho_{2}^{TQG} + \epsilon_m^2 Y\qty(\rho_{2}^{TQG}) \tag{D.6}\\
    &+ \epsilon_m\qty(1-\epsilon_m)\qty[Z\qty(\rho_{2}^{TQG}) + X\qty(\rho_{2}^{TQG})]
    \label{Eq:Measurement_Error}
\end{align*}
with $X\qty(\rho_j) \rightarrow \text{diag}\qty{C_j,D_j,A_j,B_j}$ being a bit flip and $Y\qty(\rho_j) \rightarrow \text{diag}\qty{B_j,A_j,D_j,C_j}$ being a bit and phase flip. The state distribution after entanglement swapping is then given by:
\begin{align*}
    \rho_{\text{swap}} &= A_1^{in} X\qty(\rho_2^{m}) + B_1^{in} Z\qty(\rho_2^{m}) \tag{D.7} \\
    &+ C_1^{in} \rho_2^{m} + D_1^{in} Y\qty(\rho_2^{m}) 
    \label{eq:Swapping_With_Error}
\end{align*}
with the measurement errors baked into the second path originating from $\rho_2$. 
The fidelity of the final EPR pair is then given by $\mathcal{F}=\expval{C_{\text{swap}}}$ and is lower bounded by Equation \eqref{eq:Final_EPR_fidelity}. Similarly, the background light induced error comes in linearly in this regime.

\subsection{Effect of background light}
\label{appendix:Background_Light}
Ideally, the measurement station can be thought of as a black box with a single input, corresponding to the emitted photons by the two nodes, and a single output, corresponding to the detection outcomes for the input system. However, in practice, noise can introduce additional input that influence the detection outcomes in the same way as the emitted photons. As a results, the observed detector clicks cannot be solely attributed to the original system, but may also be due to noise. This give rise to a fidelity error, while also altering $p_{\mathrm{HEG}}$ if the background level is high.

Throughout this section, we will use the following vector notation to donate the number of detector clicks:

\begin{align*}
    [c_L\, c_E \, d_L \, d_E]
    \tag{E.1}
\end{align*} 
where:
\begin{itemize}
    \item $c_L \ (c_E)$ represent the number of late (early) photons detected at one of the detectors.
    \item $d_L \ (d_E)$ represent the number of late (early) photons detected at the other detector.
\end{itemize}
 For a certain detection pattern, $[a\, b\, c\, d]$, the observed counts arise as a combination between the entangled photons and the noise:
\begin{align*}
    [a\, b\, c\, d] = [a_s\, b_s\, c_s\, d_s] + [a_n\, b_n\, c_n\, d_n]
    \label{combination}
    \tag{E.2}
\end{align*}
where the subscript $s$ and $n$ denote contribution from the entangled photons and noise, respectively. Their summation equals to the total number of detected events.

 The main sources of noise can be contributed to background light that gets coupled into the system from the channel, as well as the dark count of the detectors. With the advancement in single photon detectors, the dark counts of some detectors can reach as low as few Hz \cite{hadfield2009single}, therefore, only the background light will be studied.

The effect of background light can be understood by looking at its effect on the system through the resulting density matrix $\rho^{'}$. The density matrix $\rho^{'}$ for a certain detection pattern $[a\, b\, c\, d]$ is obtained by summing over all combinations of system, represented by the density matrix $\rho$, and background events that can lead to this outcome as follow:
\begin{align*}
    \begin{aligned}
        \rho^{'}_{\mathrm{a b c d}} = \hspace{-2.5ex} \sum_{\substack{a',b', \\ c',d'=0}}^{\substack{a,b,\\ c,d}}\hspace{-2.5ex} \, P_{c_L}(a-a')\, P_{c_E}(b-b')\, P_{d_L}(c-c') \\ P_{d_E}(d-d') {\scriptstyle \bra{a^{'}b^{'}c^{'}d^{'}} \rho \ket{a^{'}b^{'}c^{'}d^{'}}}
\end{aligned}
\label{updated_rho}
\tag{E.3}
\end{align*}
where $P_m(N)$ is the probability of getting N number of clicks on channel $m$ due to noise. Photons from background light arrive randomly distributed in time and are characterized by the average background photon rate, $B$. The average number of background photons detected within a given time window $\Delta t$ thus follows a Poisson distribution:
\begin{align*}
    P(N=n) = \frac{B^{n}\, e^{-B}}{n!}
    \label{poisson_dis}
    \tag{E.4}
\end{align*}

Plugging Eq. \eqref{poisson_dis} into Eq. \eqref{updated_rho}, the density matrix can be found as follow:

\begin{flalign}
\rho^{'}_{\mathrm{abcd}} =
    e^{-2B} \hspace{-1.5ex}\sum_{\substack{a',b', \\ c',d'=0}}^{\substack{a,b,\\ c,d}} \hspace{-.5ex}\left[\prod_{\substack{n=a,b, \\c,d}} \frac{(\frac{B}{2})^{n-n'}}{(n\hspace{-.5ex}-\hspace{-.5ex}n')!}\right] 
    {\scriptstyle \bra{a^{'}b^{'}c^{'}d^{'}} \rho \ket{a^{'}b^{'}c^{'}d^{'}}}               
    \tag{E.5}
    \label{rho}
\end{flalign}

Since background light only couples from the free-space side, after the beam splitter, the intensity is halved between the two outputs. Therefore, in the above equation, the average number of background photons (B) coupled into the detectors is divided by two.

To find $B$, we first need to find the amount of background light arriving at the telescope aperture of the ground station as follow \cite{er2005background}:
\begin{align*}
    B_r= \frac{H_b \, \Omega_{\mathrm{fov}} \, A_{\mathrm{rec}} \, \Delta\lambda \, \Delta t}{E_p}
    \label{brightness}
    \tag{E.6}
\end{align*}
where $H_{b}$ is the brightness of the sky in (W $\mathrm{m^{-2}}$ $\mathrm{\mu m}^{-1}$ $\mathrm{Sr}^{-1}$), $\Omega_{\mathrm{fov}}$ is the field of view, $A_{\mathrm{rec}}$ is the area of the receiver, $\Delta\lambda$ is the optical filter bandwidth, $\Delta t$ is the time-gate window of the detectors, and $E_p$ is the photon energy.

The value of $H_b$ depend heavily on wavelength, as well as, weather conditions. Typical values can be found in Ref. \cite{er2005background}. For $883\ \mathrm{nm}$ (wavelength for the Nd communication ion), typical values for different weather conditions can be found in Table \ref{Typical_brightness}.

\begin{table}[H]
\centering

\begin{tabular}{|c|c|}
\hline
\textbf{Conditions} & \textbf{Typical brightness}     \\
 & \textbf{$(\mathrm{W \, m^{-2} \, \mu m^{-1} \, Sr^{-1}})$}    \\
\hline
Clear daytime & $2$    \\
\hline
Full moon & $2 \times 10^{-3}$     \\
clear nighttime &  \\
\hline
New moon & $2 \times 10^{-4}$     \\
clear nighttime &  \\
\hline
Moonless & $2 \times 10^{-5}$     \\
clear nighttime &  \\
\hline
\end{tabular}
\caption{Comparison of typical brightness under different conditions}
\label{Typical_brightness}
\end{table}

In our scheme, the incoming beam will be coupled into a single mode fiber (SMF). Since background light is spatially random and fills the entire field of view, only a fraction of it will be coupled into the SMF, effectively providing a spatial filter. The coupling ratio of the photons reaching the detector can be expressed as follow:
\begin{align*}
    \eta_{\mathrm{bg}} =\frac{G_{\mathrm{SMF}}}{G_{\mathrm{rec}}} \approx \frac{\lambda^2}{A_{\mathrm{rec}} \Omega_{\mathrm{fov}}}
    \tag{E.7}
\end{align*}
Here, $G_{\mathrm{rec}}$ is the optical étendue for the receiver, and $G_{\mathrm{SMF}}$ is the optical étendue of the SMF, which can be approximated as $\lambda^2$ \cite{Betters:13}. Consequently, the number of background photons coupled into the SMF is given as follow:

\begin{align*}
    B= \eta_{bg}B_r=\frac{H_b \, \lambda^2 \, \Delta\lambda \, \Delta t}{E_p}
    \tag{E.8}
    \label{brightness_smf}
\end{align*}

Furthermore, wavelength and temporal filtering can be used to further reduce the amount of background light, represented by $\Delta\lambda$ and $\Delta t$, respectively. Typical narrow-band fiber-based optical filters have bandwidths ranging from tens of picometers to tens of nanometers depending on their type \cite{komukai2002efficient,weiss2015setup}. Even narrower bandwidth is feasible with ultra narrow-band filters achieving bandwidth of just few picometers or less \cite{maron2025few,lumeau2010ultra}. In this article, we assume a filter with a bandwidth of $0.01\ \mathrm{nm}$ to estimate the background level. For a cavity quality factor of $10^6-10^7$, the lifetime for the Nd communication ions could reach around $100\ \mathrm{ns}$ \cite{PhysRevB.108.075107,Kinos2021}. Taking that into account, the gating time for our detector is assumed to be around $200\ \mathrm{ns}$, which corresponds to a photon collection probability of $\approx 87\%$ within the gate time. This assumption ensures high collection efficiency while strongly suppressing the background light. The level of background light within the gate time $\Delta t$ for different weather conditions can be found in Table \ref{number_of_photons}:

\begin{table}[H]
\centering

\begin{tabular}{|c|c|}
\hline
\textbf{Conditions} & \textbf{B}     \\
\hline
Clear daytime & $1.38 \times 10^{-5}$    \\
\hline
Full moon & $1.38 \times 10^{-8}$     \\
clear nighttime &  \\
\hline
New moon & $1.38 \times 10^{-9}$     \\
clear nighttime &  \\
\hline
Moonless & $1.38 \times 10^{-10}$     \\
clear nighttime &  \\
\hline
\end{tabular}
\caption{Comparison of number of photons, B, coupled into the detection system under different conditions}
\label{number_of_photons}
\end{table}
\subsubsection{Background light effect on $p_{\mathrm{HEG}}$:}
An HEG trial is considered successful when we detect one early and one late photon. When background light is coupled to the system, the probability of obtaining this results, and hence $p_{\mathrm{HEG}}$, is altered. $p_{\mathrm{HEG}}$ is found by tracing over the density matrix $\rho'_{\mathrm{abcd}}$ in Eq. \eqref{rho} for the detections that give one early and one late photons as follow: 
\begin{align*}
    \begin{aligned}
     p_{\mathrm{HEG}} &= Tr(\rho{'}_{1100})+ Tr(\rho{'}_{1001}) \\ &\hspace{7
     ex} +Tr(\rho{'}_{0110}) + Tr(\rho{'}_{0011}) \\
        &=  \frac{1}{2}\eta_0^2\eta_c\,\mathcal{E}(B)
    \end{aligned} 
    \tag{E.9}
\end{align*}
where $\frac{1}{2} \eta_0^2 \eta_c$ is $p_{\mathrm{HEG}}$ when there is no background light. The factor ${1}/{2}$ accounts only for the events where one early and one late photon arrive to create a successful EPR pair \cite{Kinos2025}. And $\mathcal{E}\qty(B)$ is the effect of the background light:
\begin{align*}
    \begin{aligned}
     \mathcal{E}(B) = e^{-2B} \left( B \left( \frac{2}{\eta_0}  + \frac{2}{\eta_0\eta_c} -4 \right) + \right. \\ \left. B^2 \left( 2- \frac{2}{\eta_0} - \frac{2}{\eta_0\eta_c} +\frac{2}{\eta_0^2 \eta_c} \right) + 1 \right)
    \end{aligned} 
    \tag{E.10}
\end{align*}



\subsubsection{Background light effect on fidelity:}
Not all successfully heralded HEG trials correspond to successful entanglement, as background light may mimic one of the valid patterns while the photon entangled to one of the qubits is lost. Such events reduce the overall system fidelity as follow:


\begin{align*}
    \begin{aligned}
        \epsilon_B = 1-\frac{e^{-2B}}{4\ p_{\mathrm{HEG}}}\left[ B(\eta_0 + \eta_0\eta_c - 2\eta_0^2\,\eta_c) \right. \\ \left. -B^2\left((1-\eta_0\eta_c)(1-\eta_0)\right)   + 2\eta_0^2 \eta_c \right] \\
    \end{aligned}
    \tag{E.11}
\end{align*}








\end{document}